\documentclass{statsoc}

\usepackage[hyphens,spaces,obeyspaces]{url}
\usepackage{graphicx}
\usepackage{amssymb}
\usepackage{enumitem}

\usepackage[letterpaper]{geometry}
\usepackage{array}
\usepackage{multirow}
\usepackage{amsmath}
\usepackage{setspace}
\usepackage{esint}
\usepackage[authoryear]{natbib}

\usepackage{amsthm}
\usepackage{thmtools}\usepackage{latexsym}

\usepackage{color}
\usepackage{tikz,pgf}
\usepackage{amssymb}
\usepackage{setspace}
\usepackage{algorithmicx}
\usepackage{algorithm}
\usepackage{algpseudocode}
\usepackage[unicode=true,pdfusetitle,
 bookmarks=true,bookmarksnumbered=false,bookmarksopen=false,
 breaklinks=false,pdfborder={0 0 1},backref=section,colorlinks=true]
 {hyperref}
\hypersetup{
 citecolor=blue,linkcolor=blue,urlcolor=blue,filecolor=black}
\usepackage{breakurl}
\usepackage{xcolor}
\definecolor{col1}{RGB}{103,0,13}
\definecolor{col2}{RGB}{165,15,21}
\definecolor{col3}{RGB}{203,24,29}
\definecolor{col4}{RGB}{239,59,44}
\definecolor{col5}{RGB}{251,106,74}
\definecolor{col6}{RGB}{158,202,225}
\definecolor{col7}{RGB}{66,146,198}
\definecolor{col8}{RGB}{8,81,156}
\definecolor{col9}{RGB}{8,48,107}
\definecolor{col10}{RGB}{247,251,255}
\def\ci{\perp\!\!\!\perp}

\newcommand{\openr}{\hbox{${\rm I\kern-.2em R}$}}
\newcommand{\openn}{\hbox{${\rm I\kern-.2em N}$}}

\title{Causal inference, social networks, and chain graphs}
\author{Elizabeth L. Ogburn}
\address{Department of Biostatistics, Johns Hopkins Bloomberg School of Public
Health, Baltimore, MD, USA}
\email{eogburn@jhsph.edu}
\author{Ilya Shpitser}
\address{Department of Computer Science, Johns Hopkins University, Baltimore, MD,
USA}
\author [Ogburn {\it et al.}]{Youjin Lee}
\address{Center for Causal Inference, University of Pennsylvania, Philadelphia, PA, USA}

\begin{document}

\def\spacingset#1{\renewcommand{\baselinestretch}%
	{#1}\small\normalsize} \spacingset{1}


\begin{abstract}
Traditionally, statistical and causal inference on human subjects rely on the assumption that individuals are independently affected by treatments or exposures.  However, recently there has been increasing interest in settings, such as social networks, {where individuals may interact with one another such that} treatments may spill over from the treated individual to their social contacts and outcomes may be contagious.  Existing models proposed for causal inference using observational data from {networks of interacting individuals} have two major shortcomings. First, they often require a level of granularity in the data that is practically infeasible to collect in most settings, and second, the models are high-dimensional and often too big to fit to the available data.  In this paper we illustrate and justify a parsimonious parameterization for {network data} with interference and contagion.  Our parameterization corresponds to a particular family of graphical models known as chain graphs.  We argue that, in some settings, chain graph models approximate the marginal distribution of a snapshot of a longitudinal data generating process on interacting units. 
We illustrate the use of chain graphs for causal inference about collective decision making in social networks using data from U.S. Supreme Court decisions between 1994 and 2004 and in simulations.

\end{abstract}

\section{Introduction \label{sec:Introduction}}

Traditionally, statistical and causal inference on human subjects rely on the assumption that individuals are independently affected by treatments or exposures.  This is sometimes referred to as the \emph{no interference} assumption; it is also part of the \emph{stable unit treatment value} assumption (SUTVA).  However, recently there has been increasing interest in settings where treatments "spill over" from the treated individual to his or her social contacts, or where outcomes are contagious.  Researchers interested in causal inference have developed methods for \emph{interference} -- when one individual's treatment or exposure affects not only his own outcome but also the outcomes of his contacts \citep{aronow2012estimating,athey2016exact,Bowers2013interference,eckles2014design,forastiere2016identification,graham2010measuring,halloran2011causal,hong2006evaluating,hong2008causal,hudgens2008toward,jagadeesan2017designs,liu2014large,liu2016inverse,ogburn2014causal,rosenbaum2007interference,rubin1990application,sobel2006randomized,tchetgen2010causal,vanderweele2010direct}. Researchers interested in social networks have attempted to model the spread of contagious outcomes across network ties \citep{christakis2007spread, christakis2008collective, christakis2010social, ali2009estimating, cacioppo2009alone, lazer2010coevolution, rosenquist2010spread}, but existing methods for such modeling are either flawed \citep{cohen2008obesity, lyons2011spread, shalizi2011homophily} or limited by strong assumptions and burdensome data requirements.  One stumbling block for inference about {networks of interacting individuals}  is dimensionality: in many settings, if $n$ individuals can interfere with or transmit to one another, all $n$ outcomes are dependent, resulting in a saturated likelihood with the number of parameters growing exponentially in $n$ even before treatments and covariates are included. In this paper we illustrate a parsimonious parameterization for {such} social network data and explore when this new parameterization might be justified.  Our parameterization corresponds to a particular family of graphical models known as \emph{chain graphs}.
 Different types of chain graph models have been proposed in the literature \citep{drton2009discrete
 }.  In this paper we exclusively consider chain graph models under the Lauritzen, Wermuth, and Frydenberg (LWF) interpretation, due to the fact that these models generalize graphical models associated with both DAGs and undirected graphs, and form curved exponential families under many parameterizations \citep{lauritzen96graphical}.

\cite{lauritzen2002chain} defined causal models using chain graphs via structural equation semantics and Gibbs sampling.  \cite{shpitser2017modeling} formulated causal chain graphs using the potential outcomes semantics as a general approach to interference problems, and considered how causal mediation analysis ideas generalize to chain graph models to yield a principled approach to analyzing contagion and infectiousness effects in interference contexts.
\cite{tchetgen2017auto}, which assumed a conventional causal model for data drawn from units in a network, used chain graph Markov assumptions as a model on the observed data to make inference possible in the
\emph{full interference} setting, where the effective sample size is $1$.
\cite{sherman18dep} gave a complete identification theory for causal effects in the presence of both network dependence and unobserved confounding in latent variable chain graph models models, while \cite{bhattacharya19cinu} gave a model selection algorithm in situations where the exact structure of a network given by a chain graph is not known.  \cite{pena19unifying} described a general chain graph model which unified different types of chain graphs, allowing for different types of network relationships to be represented by different types of edges in the unified model.
\citet{lauritzen2002chain} argued against the cavalier use of chain graphs as causal models, 
but justified their use when interest is in a structured system which contains 
equilibrium distributions, as happens when outcomes
of interest represent collective behavior or collective decisions across interacting individuals. 
However, the true data-generating process where equilibrium-generating dynamics are absent are often better represented by
a directed acyclic graph (DAG) model. Causal models of DAGs 
have been used for decades to guide inference and modeling, especially for causal inference \citep{pearl2000causality}.

In this paper, we explore possible justifications for the use of chain graphs to model data on interacting units, 
where we assume a latent variable causal DAG model, but use chain graphs as a tractable approximation of the resulting observed data likelihood.  
Although chain graph models are known to be incompatible with such DAG models in general, we show in simulations that, in certain settings, the conditional independences entailed by a chain graph model may approximate those from a DAG model with certain properties. 
In addition, we apply the chain graph approach to data on Supreme Court cases, where the underlying data generating process may be viewed as containing equilibrium dynamics, as opinions about a case get gradually formed.
Using chain graph models for causal inference with social network data extends a number of papers that have used undirected graph models, or Markov random fields, to model social interactions, including the "sociophysics" literature on Ising models for collective decision making.

The rest of the paper is organized as follows: Section \ref{sec:graphtheory} reviews required concepts, definitions, and notation pertaining to undirected, directed, and chain graph models; readers familiar with graphical models may be able to skip this section. In Section \ref{sec:history}, we describe previous work using DAGs for causal inference with social network data and in settings with contagion and interference, and previous work using undirected graphical models to study collective problem solving.  We propose conditions under which chain graphs may approximate a true underlying DAG model for social network data and illustrate the relationship between the DAG model and a chain graph approximation in simulations.  We explain how chain graphs can be used to analyze data with contagion and interference when DAG models are intractable.  These chain graph models extend previously proposed models for collective problem solving in important and useful ways. In Section \ref{sec:data} we analyze data on Supreme Court decisions in order to illustrate how chain graph models can be used to estimate causal effects on collective outcomes.  Only group-level exposure variables are available in the real data, so in Section \ref{ssec:supreme_sim} we simulate data based on the Supreme Court decision data but with individual-level exposures in order to illustrate the estimation of individual-level causal effects using chain graph models. Section \ref{sec:conclusion} concludes the paper. 

\section{Graphs and Graphical Models} \label{sec:graphtheory}

Graphical models use graphs--collections of vertices, representing random variables, and edges representing relations between pairs of vertices--to concisely represent conditional independences that hold among the random variables.  At their most general, the graphical models we will consider in this paper are represented by mixed graphs containing directed ($\to$), and undirected
($-$) edges, such that at most one edge connects two vertices. This section contains a brief, informal review of graphical models; a more comprehensive review is in the Appendix.

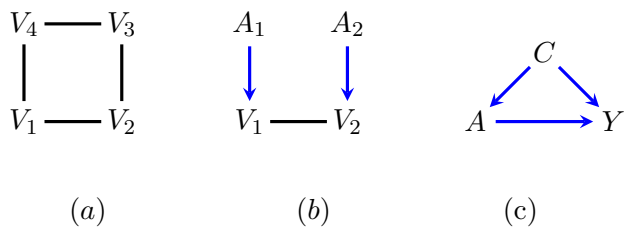
\begin{figure}
\begin{centering}
\begin{tikzpicture}[>=stealth, node distance=1.3cm]
    \tikzstyle{format} = [ very thick, circle, minimum size=5.0mm,
	inner sep=0pt]
    \tikzstyle{square} = [very thick, rectangle, draw]

	\begin{scope}[xshift=0.0cm]
		\path[->, very thick]
			node[format] (v1) {$V_1$}
			node[format, right of=v1] (v2) {$V_2$}
			node[format, above of=v2] (v3) {$V_3$}
			node[format, left of=v3] (v4) {$V_4$}
			
			(v1) edge[-] (v2)
			(v2) edge[-] (v3)
			(v3) edge[-] (v4)
			(v1) edge[-] (v4)
			
			node[below of=v2, xshift=-0.45cm, yshift=0.1cm] (l) {$(a)$}

			;
	\end{scope}
	
		\begin{scope}[xshift=3.0cm]
		\path[->, very thick]
			node[format] (v1) {$V_1$}
			node[format, right of=v1] (v2) {$V_2$}
			node[format, above of=v2] (v3) {$A_2$}
			node[format, left of=v3] (v4) {$A_1$}
			
			(v1) edge[-] (v2)
			(v3) edge[->, blue] (v2)
			(v4) edge[->, blue] (v1)
			
			node[below of=v2, xshift=-0.45cm, yshift=0.1cm] (l) {$(b)$}

			;
	\end{scope}
        \begin{scope}[xshift=6.0cm]
                \path[very thick, ->]
                        node[format] (a) {$A$}
                        node[format, above right of=a] (u) {$C$}
                        node[format, below right of=u] (y) {$Y$}

                        (a) edge[blue] (y)
                        (u) edge[blue] (a)
                        (u) edge[blue] (y)

		node[below of=a, xshift=0.60cm, yshift=0.1cm] (l) {(c)}
                    
                ;
        \end{scope}
\end{tikzpicture} 
\par\end{centering}
\caption{(a) A simple undirected graph.  (b) The simplest chain graph with an independence model not representable as either a DAG or an undirected graph.
(c) A causal graph representing observed confounding of the treatment $A$ and outcome $Y$ by a set of covariates $C$.}
\label{fig:graph} 
\end{figure}

 A graph with only undirected edges an \emph{undirected graph} (UG), as in Figure \ref{fig:graph}(a). Such graphs are often used to model networks of connected individuals; the absence of an edge between two nodes indicates, roughly, that the nodes are independent conditional on other intermediate nodes. When we analyze the U.S. Supreme Court data, we will use an undirected graphical model to represent the pairwise relationships among the justices, and the absence of an edge will indicate that there is no evidence of a direct connection between the two justices beyond what can be explained by individual level variables or by other pairwise connections in the network of justices.
 
 A graph with only directed edges is called a \emph{directed acyclic graph} (DAG), as in Figure \ref{fig:graph}(c).  Such graphs underpin most causal inference methods, with the absence of an edge between two nodes indicating, roughly, that there is no causal effect of one on the other that is not mediated by other nodes on the graph.  Figure \ref{fig:graph}(c) depicts a setting in which $A$ causes $Y$ and $C$ is a common cause, or a confounder, of the $A$-$Y$ relationship. 

A mixed graph containing both directed and undirected edges with no partially directed cycles is called a \emph{chain graph} (CG) \citep{lauritzen1989graphical,frydenberg1990chain}. A simple example is depicted in Figure \ref{fig:graph}(b). As we describe in detail below, graphs of this type may be used to represent systems with interacting units.
 A chain graph forms a natural generalization of both statistical models associated with DAGs, and associated with UGs.
However, the combination of directed and undirected edges can result in conditional independences that would not be familiar or intuitive to practitioners familiar with DAG and UG models.
For example, in Figure~\ref{fig:graph} (b),  $A_1$ is independent of $ V_2$ given $V_1$ and $A_2$, as one might expect, but $A_1$ is \emph{not} independent of  $V_2$ given only $V_1$. 
The factorization that defines chain graph models and implies the constraints above in the special case of the model shown in Figure~\ref{fig:graph} (b) is given in the Appendix.

DAGs and CGs have both been used to define statistical and causal models.
Statistical graphical models associate the observed data distribution $p({\bf V})$ with a graph where vertices are associated with random variables in ${\bf V}$.
Graphical models are often defined by means of a factorization, where $p({\bf V})$ may be written as a product of smaller factors, with a rule for obtaining these factors given by the graph.

Causal inference from observational data is concerned with making inferences on \emph{counterfactual} or \emph{potential outcome} random variables of the form $Y(a)$ from the observed data distribution $p({\bf V})$.  Here, $Y(a)$ is taken to mean ``the variable $Y$ had we intervened on $A$, possibly contrary to fact, to set it to $a$.''  Causal parameters of primary interest are usually low dimensional summaries or contrasts obtained from counterfactual distributions, rather than  counterfactual distributions themselves.  For example, the \emph{average causal effect (ACE)}, is defined as the mean contrast $\mathbb{E}[Y(a)] - \mathbb{E}[Y(a')]$.
Causal graphical models
define a link between counterfactual targets of inference, such as the ACE, and the observed data distribution.
Causal graphical models 
imply a factorization, and thus a statistical graphical model, on the observed data distribution.  The link between this distribution and counterfactual parameters is provided by means of a modified factorization of the observed data distribution.  In the case of causal models of a DAG, this modified factorization is known as the g-formula \citep{robins86new}.  In the case of causal models of a chain graph, this modified factorization is described in the Appendix, and originally in \citep{lauritzen2002chain}.
Causal DAG models are powerful tools for causal inference using observational data and have gained widespread use in epidemiology, social sciences, and other fields, because they can be used to clearly display sources of bias such as confounding, and can be used to derive identification theory for many counterfactual targets of inference in complex multivariate causal systems.

 Chain graphs allow both directed and undirected edges, and can be used to define 
  both statistical and causal graphical models that combine features of both undirected graphs and DAGs \citep{lauritzen96graphical}.  For simplicity, in this paper we will restrict ourselves to chain graphs with an undirected component or block ${\bf Y}$, representing outcomes associated with nodes in a network, and with exposures ${\bf A}$ having directed edges into ${\bf Y}$.  
We will only consider interventions on ${\bf A}$--here we will not consider interventions on nodes in the undirected component of outcomes.  A general treatment of chain graph models may be found in \citep{lauritzen96graphical, lauritzen2002chain, sherman18dep}.


\section{Graphical models for social interactions} \label{sec:history} 

DAG models are assumed either implicitly or explicitly in almost all existing methods for learning about social interactions, social influence, interference, contagion, and other causal effects from network data.  Causal DAG models, or the mathematically equivalent causal structural equation models, correspond to a mechanistic view of the (macroscopic) world, which is espoused by most researchers across many disciplines and underpins almost all approaches to learning about causal effects from data. 
 \citet{ogburn2014causal} provides an overview of the use of DAGs to represent interference and contagion.  New methods for learning about causal effects from social network data rely on assumptions that are consistent with DAG models,
 explicitly in the case of methods for observational data proposed by \citet{van2014causal} and \citet{ogburn2017causal} and implicitly in many of the methods based on randomized experiments (e.g. \citealp{aronow2012estimating, athey2016exact,Bowers2013interference, choi2014estimation, eckles2014design,forastiere2016identification, graham2010measuring, hong2006evaluating, hong2008causal, hudgens2008toward, jagadeesan2017designs, liu2014large, liu2016inverse, rosenbaum2007interference, rubin1990application,sobel2006randomized,tchetgen2010causal,vanderweele2010direct}).  However, as we will show in the next section and as has been acknowledged by some of the aforementioned researchers, DAGs in these settings can run into difficulties in practice. 

Undirected graph models have also been used to model social networks (see, e.g., \citealp{west2014exploiting, domingos2001mining, ahmed2009recovering, kindermann1980relation}), and a niche literature uses a particular class of undirected graph models, namely \emph{Ising models}, to model the collective behavior of individual actors.  
The Ising model was originally developed by physicists to model spin-states (up/down)
of atoms of a metal arranged on a lattice \citep{ising25beitrag}; it is meant to represent the state of a physical system
at a particular temperature.  As temperature decreases but remains strictly
greater than absolute zero, the system may transition from
a relatively disordered state to a state where most spins are either up
or down. Such a phenomenon is called a \emph{phase transition}. Phase
transitions do not occur for Ising models\textbf{ }in a one dimensional
lattice, but they occur in all higher dimensional grids, starting
with dimension $2$ \citep{peierls1936ising}. A literature has developed that uses the Ising model to represent
the forming
of consensus in groups of interacting individuals,
each holding one of a pair of possible opinions, e.g. for or against
a proposition (\citealt{galam1982sociophysics,galam1997rational,sznajd2000opinion};
see also \citealp{galam2008sociophysics} and references therein).  A phase transition occurs if individuals converge to a consensus.  The rectangular lattice underlying the model
restricts each individual to interacting with a fixed number of other
individuals, depending on the dimension of the lattice. Despite the
fact that these models have not been fit, validated, or tested against
real data, and despite the fact that social networks are usually nowhere close to being rectangular lattices, sociophysicists have made strong and empirically verifiable
claims about social phenomena using these models (e.g.  \citealp{galam2008sociophysics}).

The Ising model is a special case
of a log-linear model defined on an undirected graph.  
Chain graph models, 
which include undirected graphs as a special case, generalize
Ising models for human behavior in several directions: to arbitrary network structure, rather than lattices; to facilitate statistical inference and model fitting using real data; to introduce the idea of treatments with causal effects on nodes; and to clarify the uses and limitations of such models for human behavior.  These chain graph models can also be seen to extend other undirected models for social networks.

\begin{figure}
\begin{centering}
\begin{tikzpicture}[>=stealth, node distance=1.3cm]
    \tikzstyle{format} = [ very thick, circle, minimum size=5.0mm,
	inner sep=0pt]
    \tikzstyle{square} = [very thick, rectangle, draw]

                \begin{scope}[xshift=4cm]
                \path[very thick, ->]
                        node[format] (a1) {$A_1$}
                        node[format, below of=a1] (a2) {$A_2$}
                        node[format, below of=a2] (a3) {$A_3$}

                        node[format, gray, right of=a1] (y11) {$Y^1_1$}
                        node[format, gray, right of=a2] (y12) {$Y^1_2$}
                        node[format, gray, right of=a3] (y13) {$Y^1_3$}

                        node[format, gray, right of=y11] (y21) {$Y^2_1$}
                        node[format, gray, right of=y12] (y22) {$Y^2_2$}
                        node[format, gray, right of=y13] (y23) {$Y^2_3$}

                        node[format, right of=y21] (d1) {$\ldots$}
                        node[format, right of=y22] (d2) {$\ldots$}
                        node[format, right of=y23] (d3) {$\ldots$}

                        node[format, right of=d1] (y31) {$Y^T_1$}
                        node[format, right of=d2] (y32) {$Y^T_2$}
                        node[format, right of=d3] (y33) {$Y^T_3$}

			(a1) edge[blue] (y11)
			(a1) edge[blue, bend left=25] (y21)
			(a1) edge[blue, bend left=25] (y31)

			(a2) edge[blue] (y12)
			(a2) edge[blue, bend left=25] (y22)
			(a2) edge[blue, bend left=25] (y32)

			(a3) edge[blue] (y13)
			(a3) edge[blue, bend left=25] (y23)
			(a3) edge[blue, bend left=25] (y33)
			
			(y11) edge[blue] (y21)
			(y12) edge[blue] (y22)
			(y13) edge[blue] (y23)

			(d1) edge[blue] (y31)
			(d2) edge[blue] (y32)
			(d3) edge[blue] (y33)
			
			(y21) edge[blue] (d1)
			(y22) edge[blue] (d2)
			(y23) edge[blue] (d3)

			(y11) edge[blue] (y22)
			(y13) edge[blue] (y22)
			(y12) edge[blue] (y21)
			(y12) edge[blue] (y23)

			(y21) edge[blue] (d2)
			(y23) edge[blue] (d2)
			(y22) edge[blue] (d1)
			(y22) edge[blue] (d3)

			(d1) edge[blue] (y32)
			(d3) edge[blue] (y32)
			(d2) edge[blue] (y31)
			(d2) edge[blue] (y33)

			node[below of=y23, xshift=0.0cm, yshift=0.3cm] (l) {(a)}                    
                ;
        \end{scope}

        \begin{scope}[xshift=11.0cm]
                \path[very thick, ->]
                        node[format] (a1) {$A_1$}
                        node[format, below of=a1] (a2) {$A_2$}
                        node[format, below of=a2] (a3) {$A_3$}

                        node[format, right of=a1] (y1) {$Y^T_1$}
                        node[format, right of=a2] (y2) {$Y^T_2$}
                        node[format, right of=a3] (y3) {$Y^T_3$}

			(a1) edge[blue] (y1)
			(a2) edge[blue] (y2)
			(a3) edge[blue] (y3)
			
			(a1) edge[blue] (y2)
			(a2) edge[blue] (y1)
			(a2) edge[blue] (y3)
			(a3) edge[blue] (y2)
			
			(y1) edge[<->, red, bend left=50] (y2)
			(y2) edge[<->, red, bend left=50] (y3)
			(y1) edge[<->, red, bend left=65] (y3)

			node[below of=a3, xshift=0.45cm, yshift=0.3cm] (l) {(b)}
                ;
        \end{scope}
        \begin{scope}[xshift=15.0cm]
                \path[very thick, ->]
                        node[format] (a1) {$A_1$}
                        node[format, below of=a1] (a2) {$A_2$}
                        node[format, below of=a2] (a3) {$A_3$}

                        node[format, right of=a1] (y1) {$Y^T_1$}
                        node[format, right of=a2] (y2) {$Y^T_2$}
                        node[format, right of=a3] (y3) {$Y^T_3$}

			(a1) edge[blue] (y1)
			(a2) edge[blue] (y2)
			(a3) edge[blue] (y3)

			(y1) edge[-] (y2)
			(y2) edge[-] (y3)

			node[below of=a3, xshift=0.45cm, yshift=0.3cm] (l) {(c)}                    
                ;
        \end{scope}

\end{tikzpicture} 
\par\end{centering}
\caption{
(a) Causal DAG representing opinion formation among peers.  $A_i$ represents interventions meant to influence subject $i$, $Y_i^k$ is the $i$th subject's opinion at time $k$.
(b) A latent projection of the model in (a) onto variables $A_1,A_2,A_3,Y^T_1,Y^T_2,Y^T_3$, representing the distribution of opinion in (a) at time $T$, before equilibrium is reached. The red bidirected arrows represent the fact that the outcomes at intermediate time points are unmeasured common causes of the observed outcomes.
(c) A chain graph model that approximates the distribution of opinion in (a) at time $T$ under certain data generating processes.
}
\label{fig:dag-temporal} 
\end{figure}
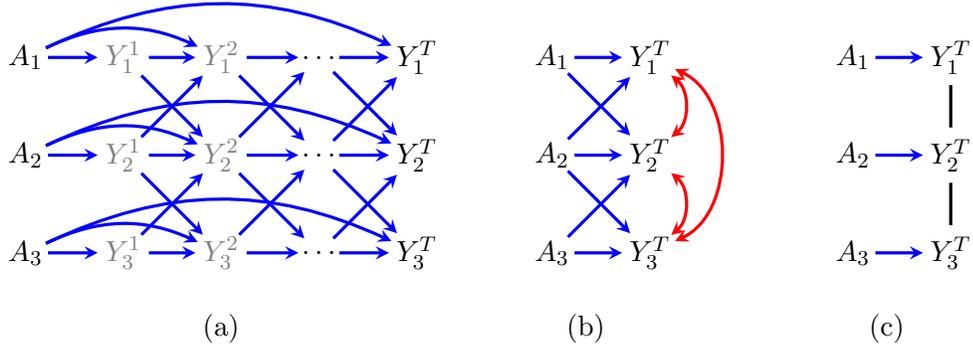

Consider a social network of $n$ individuals, or nodes.  Node $i$ is associated with a treatment or exposure, $A_{i}$, an outcome $Y_{i}$, and possibly covariates.  For example, $Y$ could represent opinions and $A$ advertising campaigns; $Y$ could represent behavior and $A$ encouragement interventions, or $Y$ could represent an infectious disease and $A$ vaccination. In our analysis of Supreme Court decisions below, $Y_i$ is a binary variable representing whether Justice $i$'s decision was liberal or conservative, and $A$, which simultaneously treats all of the Justices, is an indicator of the issue area of the case.  We specifically consider chain graphs in which the set of outcomes $Y$ form a single, undirected block, while treatments and covariates are represented by vertices with directed edges into $Y$-vertices.  A simple example is depicted in Figure \ref{fig:dag-temporal} (c).

When individual's beliefs or opinions undergo phase transitions to orderly states, e.g. when there is external pressure to reach a unanimous consensus, or when it can be argued that the distribution of individual's behaviors, beliefs, opinions, or other outcomes attains an equilibrium across network ties, then a chain graph may be the correct model for the joint distributions of outcomes across a network and interventions on those outcomes.  For example, in the Supreme Court data, outcomes represent decisions made under time constraints and with pressure for the nine justices to reach a unanimous decision; these may indeed be in equilibrium. More common in the existing literature are settings in which DAG models would be the most appropriate class of models, but they are not tractable given reasonable constraints on data collection.  


Figure \ref{fig:dag-temporal} (a) depicts a DAG model for a three-node network in which individuals 1 and 2 exhibit contagion, as do 2 and 3, where contagion is any causal effect of one individual's outcome at a particular time on their social contacts' future outcomes--or an arrow from $Y_{i}^{t}$ to $Y_{j}^{s}$ for $t<s$.  In opinion formation, like the Supreme Court decision data, contagion could be the influence that one Justice's way of thinking about a case can have on other Justice through conversation or debate.  The pairs of individuals who influence one another often correspond to pairs of individuals with ties in an observed social network. Figure \ref{fig:dag-temporal} (a) does not include any direct interference; this would be a causal effect of one individual's treatment on another's outcome, that is an arrow from $A_i$ to $Y_{j}^t$.  

In order for this DAG to be valid, the units of time captured must be small enough that any influence
passing from individual 1 to 3 through 2 cannot occur in fewer than 2 time steps
\citep{ogburn2014causal}. This will be the case if influence can
only occur during discrete interactions such as in-person or online
encounters, and the unit of time is chosen to be the minimum time
between encounters. This DAG model encodes several conditional independences,
and if we are able to observe the outcome for all agents at all time steps,
inference under these sorts of models may be possible \citep{ogburn2017causal}.


However, in most practical applications, with the exception of online social networks, it is only be possible to observe the outcome at one or a few time points. If data are generated according to the DAG in Figure \ref{fig:dag-temporal} (a) but the outcome is observed at only one time point (at which the outcome is not in a chain graph equilibrium), then the resulting model is represented by a mixed graph representing the \emph{latent projection} \citep{verma90equiv} of all of the variables in Figure \ref{fig:dag-temporal} (a) onto the subset of those variables that are actually observed, with \emph{bidirected} edges representing the presence of one or more hidden common causes.
A general construction algorithm for these latent projection mixed graphs is given by \citep{pearl09causality}, and the result for Figure \ref{fig:dag-temporal} (a) is shown in Figure \ref{fig:dag-temporal} (b).  

Collecting or accessing the detailed temporal data required to use the models like Figure \ref{fig:dag-temporal} (a) is often impractical or impossible, but the saturated model for the marginal in Figure \ref{fig:dag-temporal} (b) quickly becomes unwieldy, as the number of parameters required to estimate and use the model grows exponentially with the number of nodes: the latent projection graph will generally not be sparse,even if the underlying social network governing opinion formation is.  To see this, note that after a single time step,
an individual only influences neighboring individuals, but after two time steps, also neighbors of neighbors.  In the three-person network represented by Figure \ref{fig:dag-temporal}, this is enough to render the latent projection of Fig \ref{fig:dag-temporal} (b) fully saturated, with no conditional independences.  After
many time steps, the individual's influence would have time to reach most of the social network.  This implies that
any two outcomes at time $t$, for a large enough $t$, will be related via a chain of hidden common causes, even
if the corresponding individuals are far from each other in the social network.  To represent these chains of hidden common
causes, the latent projection graph would contain a clique of bidirected edges encompassing opinions of everyone in the network.
Given a mixed graph containing such a clique, the number of parameters needed to specify the appropriate mixed graph likelihood will be exponential in the size of the clique in general \citep{evans14markovian}.
These limitations are reflected in the literature, which posits DAG models (either explicitly or implicitly) but rarely includes applications to real data. 

Unlike the model in Figure \ref{fig:dag-temporal} (b), the chain graph model represented by Figure \ref{fig:dag-temporal} (c) is not saturated, and generally has a likelihood with the number of parameters polynomial in the size of the graph.

In general, local interactions between units leading to a causal model of the type shown in Fig.~\ref{fig:dag-temporal} (a) cannot correctly be represented by undirected edges of a chain graph.  Indeed, most seemingly reasonable uses of such edges lead to inconsistencies or Markov properties not represented by chain graphs, as discussed extensively in \cite{lauritzen2002chain}.
Nevertheless, a chain graph model that is as sparse as the underlying social network may serve as a good approximation of the underlying latent variable causal model, even if the likelihood corresponding to its latent projection representation is intractable.

\begin{figure}
\begin{centering}
\begin{tikzpicture}[>=stealth, node distance=1.1cm]
    \tikzstyle{format} = [ very thick, circle, minimum size=5.0mm,
	inner sep=0pt]
    \tikzstyle{square} = [very thick, rectangle, draw]

                \begin{scope}[xshift=5cm]
                \path[very thick, ->]
                        node[square] (a1) {$A_1$}
                        node[format, below of=a1] (a2) {$A_2$}
                        node[square, below of=a2] (a3) {$A_3$}

                        node[format, gray, right of=a1] (y11) {$Y^1_1$}
                        node[format, gray, right of=a2] (y12) {$Y^1_2$}
                        node[format, gray, right of=a3] (y13) {$Y^1_3$}

                        node[format, gray, right of=y11] (y21) {$\ldots$}
                        node[format, gray, right of=y12] (y22) {$\ldots$}
                        node[format, gray, right of=y13] (y23) {$\ldots$}

                        node[format, right of=y21] (d1) {$Y^{T-1}_1$}
                        node[format, right of=y22] (d2) {$Y^{T-1}_2$}
                        node[format, right of=y23] (d3) {$Y^{T-1}_3$}

                        node[format, right of=d1] (y31) {$Y^T_1$}
                        node[square, right of=d2] (y32) {$Y^T_2$}
                        node[format, right of=d3] (y33) {$Y^T_3$}

			(a1) edge[blue] (y11)
			(a1) edge[blue, bend left=25] (y21)
			(a1) edge[blue, bend left=25] (y31)

			(a2) edge[blue] (y12)
			(a2) edge[blue, bend left=25] (y22)
			(a2) edge[blue, bend left=25] (y32)

			(a3) edge[blue] (y13)
			(a3) edge[blue, bend left=25] (y23)
			(a3) edge[blue, bend left=25] (y33)
			
			(y11) edge[blue] (y21)
			(y12) edge[blue] (y22)
			(y13) edge[blue] (y23)

			(d1) edge[red] (y31)
			(d2) edge[blue] (y32)
			(d3) edge[red] (y33)
			
			(y21) edge[blue] (d1)
			(y22) edge[blue] (d2)
			(y23) edge[blue] (d3)

			(y11) edge[blue] (y22)
			(y13) edge[blue] (y22)
			(y12) edge[blue] (y21)
			(y12) edge[blue] (y23)

			(y21) edge[blue] (d2)
			(y23) edge[blue] (d2)
			(y22) edge[red] (d1)
			(y22) edge[red] (d3)

			(d1) edge[blue] (y32)
			(d3) edge[blue] (y32)
			(d2) edge[blue] (y31)
			(d2) edge[blue] (y33)

			node[below of=y23, xshift=0.0cm, yshift=0.3cm] (l) {(a)}                    
                ;
        \end{scope}
        
            \begin{scope}[xshift=12cm]
                \path[very thick, ->]
                        node[square] (a1) {$A_1$}
                        node[format, below of=a1] (a2) {$A_2$}
                        node[square, below of=a2] (a3) {$A_3$}

                        node[format, gray, right of=a1] (y11) {$Y^1_1$}
                        node[format, gray, right of=a2] (y12) {$Y^1_2$}
                        node[format, gray, right of=a3] (y13) {$Y^1_3$}

                        node[format, gray, right of=y11] (y21) {$\ldots$}
                        node[format, gray, right of=y12] (y22) {$\ldots$}
                        node[format, gray, right of=y13] (y23) {$\ldots$}

                        node[format, right of=y21] (d1) {$Y^{T-1}_1$}
                        node[format, right of=y22] (d2) {$Y^{T-1}_2$}
                        node[format, right of=y23] (d3) {$Y^{T-1}_3$}

                        node[format, right of=d1] (y31) {$Y^T_1$}
                        node[square, right of=d2] (y32) {$Y^T_2$}
                        node[format, right of=d3] (y33) {$Y^T_3$}

			(a1) edge[blue] (y11)
			(a1) edge[blue, bend left=25] (y21)
			(a1) edge[blue, bend left=25] (y31)

			(a2) edge[blue] (y12)
			(a2) edge[blue, bend left=25] (y22)
			(a2) edge[blue, bend left=25] (y32)

			(a3) edge[blue] (y13)
			(a3) edge[blue, bend left=25] (y23)
			(a3) edge[blue, bend left=25] (y33)
			
			(y11) edge[blue] (y21)
			(y12) edge[blue] (y22)
			(y13) edge[blue] (y23)

			(d1) edge[red] (y31)
			(d2) edge[blue] (y32)
			(d3) edge[red] (y33)
			
			(y21) edge[blue] (d1)
			(y22) edge[blue] (d2)
			(y23) edge[blue] (d3)

			(y11) edge[blue] (y22)
			(y13) edge[blue] (y22)
			(y12) edge[blue] (y21)
			(y12) edge[blue] (y23)

			(y21) edge[blue] (d2)
			(y23) edge[blue] (d2)
			(y22) edge[blue] (d1)
			(y22) edge[blue] (d3)

			(d1) edge[red] (y32)
			(d3) edge[red] (y32)
			(d2) edge[blue] (y31)
			(d2) edge[blue] (y33)

			node[below of=y23, xshift=0.0cm, yshift=0.3cm] (l) {(b)}                    
                ;
        \end{scope}

\end{tikzpicture} 
\par\end{centering}
\caption{
Paths that connect $Y_{1}^{T}$ and $Y_{3}^{T}$ even when conditioning on $Y_{2}^{T}$ and $A_{1}$ and/or $A_{3}$.  Boxes indicate variables that are conditioned on.
}
\label{fig:d-sep} 
\end{figure}
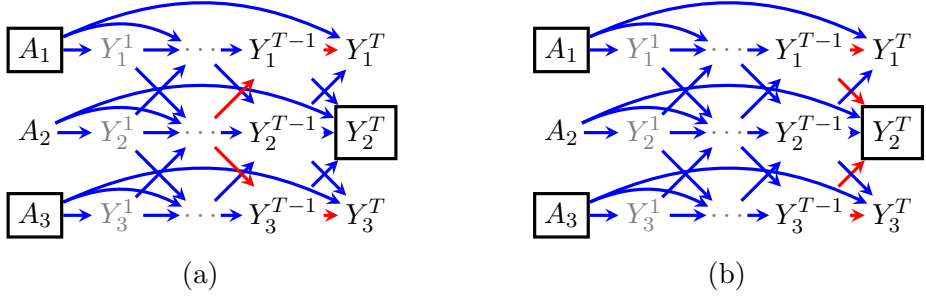

Consider chain graph models like the one in Figure \ref{fig:dag-temporal} (c), but with arbitrary undirected components corresponding to outcomes observed on social network nodes.  These models imply that each node's outcome is independent of its non-neighbors' outcomes conditional on its neighbors' outcomes and on any treatments or covariates with arrows pointing into the node.  For chain graphs like Figure \ref{fig:dag-temporal} (c), with a single treatment for each node, the conditional independences implied by the graph are of the form $Y_{i}^{T}{\ci} A_j, Y_{j}^{t}\mid A_i,\left\{ Y_{l}^{T},\forall l\mbox{ adjacent to }i\right\}$.  These conditional independences fail to hold in the corresponding DAG models due to two different types of paths, depicted in red in Figure \ref{fig:d-sep}.  

Paths like the one in Figure \ref{fig:d-sep} (a) represent the fact that the past outcomes of mutual connections affect both $Y_{i}^{T}$ and $Y_{j}^{T}$; this is just one of many such paths.  All of these paths can be blocked by conditioning on $\left\{ Y_{l}^{t},\mbox{ for all } l\mbox{ adjacent to }i \mbox{ and for }1 \leq t \leq T-1\right\}$ \citep{ogburn2017vaccines}. If the outcome evolves slowly over time, $Y_{l}^{t}$ and $Y_{l}^{T}$ will be highly correlated and conditioning on $\left\{ Y_{l}^{T},\mbox{ for all } l\mbox{ adjacent to }i \right\}$ will mostly block these paths. We expect the paths through $Y_{l}^{t}$ to be weaker for smaller $t$ than for $t$ close to $T$. If paths through $Y_{l}^{t}$ are weaker for earlier times $t$, then the relationship between $Y_{l}^{t}$ and $Y_{l}^{T}$ can also weaken for decreasing $t$ -- as long as it remains strong enough to allow conditioning on $Y_{l}^{T}$ to approximately block paths through $Y_{l}^{t}$.   

However, conditioning on $\left\{ Y_{l}^{T},\mbox{ for all } l\mbox{ adjacent to }i \right\}$ opens paths through colliders, like the one depicted in Figure \ref{fig:d-sep} (b).  M-shaped collider paths like these are known to often induce weak dependence  \citep{greenland2003quantifying}, and the magnitude can be bounded more precisely using knowledge of the partial correlation structure of the variables along the path \citep{chaudhuri2002using}. Informally, if the dependence of  $Y_{l}^{T}$ on $Y_{l}^{T-1}$ is stronger than that of $Y_{l}^{T}$ on $Y_{i}^{T-1}$ and $Y_{j}^{T-1}$, as it will be if the outcome evolves slowly over time, then the dependence induced by paths through colliders may be negligible. 

Although chain graph models exist in which the relationships along undirected edges are not symmetric, we found in simulations that DAGs with symmetric relationships for connected pairs of individuals were better approximated by chain graphs. It might be reasonable to assume this kind of symmetry if, for example, the outcome is a behavior or belief and the subjects are peers with no imbalance of power or influence, or if the outcome is an infectious disease and the subjects have similar underlying health and susceptibility statuses.




Based on the arguments above, we make the following conjecture.
\emph{Assume data on interacting units are generated from a process given by a temporal DAG model containing baseline factors, exposures, and unit outcomes evolving over time, where 
(1) there is contagion (a causal link from a network neighbor's outcome at a previous time point to the unit's present outcome) but no direct interference (causal links between neighboring units at the same time point), 
(2) outcomes evolve slowly over time, and (3) causal relationships are symmetric between pairs of units 
that share a network tie.
Then a marginal distribution containing a set of baseline factors, exposures, and outcomes at the final observed time point generated from such a process is well approximated by 
a chain graph model that links outcomes of units that share a network tie by an undirected edge while maintaining directed edges between variables within every unit.}

Formal results, for example quantifying the distance between a chain graph approximation and a true DAG model, are beyond the scope of this paper and may not be possible in full generality.  However, we verified our conjecture in simulations. We simulated ten random nine-node social networks with edge probability $p=0.3$ for every pair of nodes. For each random network, we generated outcomes for the nodes 1000 times according to a DAG model like the one in Figure \ref{fig:dag-temporal} (a), with symmetric causal effects for every edge in the underlying network.  For all nonadjacent pairs $(i,j)$, we tested three (conditional) independence hypotheses:
\begin{enumerate} [topsep=3pt,itemsep=0ex,partopsep=1ex,parsep=1ex]
\item the null hypothesis of marginal independence $Y_{i}^{t}{\ci}Y_{m}^{t}$,
\item the null hypothesis of conditional independence $Y_{i}^{t}{\ci}Y_{m}^{t}\mid \left\{ Y_{l}^{t},\forall l\mbox{ adjacent to }i\right\}$, and 
\item the null hypothesis of conditional independence $Y_{i}^{t}{\ci}Y_{m}^{t}\mid A_{i}, \left\{ Y_{l}^{t},\forall l\mbox{ adjacent to }i\right\}$.
\end{enumerate}
The conditional independences in (b) and (c) are implied by the chain graph model (for the simple case of 3 units, the associated chain graph is shown in Figure~\ref{fig:dag-temporal} (c)) but do not hold in the true model  (for the simple case of 3 units, the associated hidden variable DAG is shown in Fig.~\ref{fig:dag-temporal} (a)). We also tested (a) to ensure that we had not inadvertently generated data with such weak dependences between individuals that (b) and (c) would hold pathologically. We found that the conditional independence nulls were rejected at close to the nominal rate of 5\% expected under the null. In contrast, the marginal independence null was rejected more frequently, suggesting that conditioning on neighbors' outcomes may recover approximate independence under at least some data generating processes, and that the chain graph model may in those cases be a reasonable parsimonious approximation to the true underlying conditional independences. In another set of simulations with three agents we found that a chain graph model was able to approximately estimate causal effects from data generated under a DAG model. For details and full results see the Supplementary Materials.  

In the next sections we illustrate the use of chain graph models for causal inference about social interactions in real and simulated data on U.S. Supreme Court decisions. 

\section{Using chain graph models to analyze U.S. Supreme Court decisions} \label{sec:data}

The U.S. Supreme Court is comprised of nine justices, one of whom is the Chief Justice, tasked with presiding over oral arguments, serving as the spokesperson for the court, and other administrative roles. After a case is heard by the Supreme Court, the justices discuss and decide the case over a period of several weeks or months.  The final outcome is decided by majority vote; the majority and, when the decision is not unanimous, the minority write opinions justifying their decisions. The oral and written arguments presented to the court and the judicial opinions are public resources; however, we have no access to the debates and discussions that lead the justices to their decisions.  This precludes the use of a DAG model for the evolution of justices' opinions over time, but is amenable to a chain graph model with $Y_i$ defined as Justice $i$'s final opinion.  Indeed, such a chain graph model may be appropriate here not just as an approximation.  This is an example of collective opinion formation with pressure for unanimity, and it may attain a degenerate equilibrium in which each justice is unwilling or unable to be swayed from a fixed decision, even with further discussions.  In what follows we are agnostic as to whether the data were truly generated under a chain graph  model representing opinion formation as an equilibrium process or a DAG model; we simply assume that the chain graph model we use is a good approximation for the observed data distribution generated by the true data generating process.  However, we also give the caveat that this data analysis is meant to serve as an example illustrating the use of chain graph models for simultaneous outcomes across network nodes, rather than to draw substantive conclusions about the operations of the U.S. Supreme Court.

Data on all Supreme Court decisions since 1946, along with rich information on the nature of the cases and the opinions, is maintained by Washington University Law School's Supreme Court Database (\url{http://scdb.wustl.edu/data.php}). We used the subset of these data corresponding to the Second Rehnquist Court, a period of ten years (1994-2004) during which the same nine justices served together: William Rehnquist (Chief Justice), John Paul Stevens, Sandra Day O'Connor, Antonin Scalia, Anthony Kennedy, David Souter, Clarence Thomas, Ruth Bader Ginsburg, and Stephen Breyer.  Over these ten years the court decided 893 cases; each case serves as a datapoint for our chain graph model.

\begin{table}
	\caption{ \label{tab:case} The number of cases decided during 1994 - 2004. No cases were identified for interstate relations, miscellaneous, or private action}	
	\centering
	\fbox{%
		\begin{tabular}{*{5}{l}}
			\hline
			Issue area & Criminal procedure & Civil rights & First amendment & Due process \\
			Number of cases & 231 & 161 & 59 & 43 \\
			\hline
			Issue area &  Privacy & Attorneys & Unions & Economic Activity \\
			Number of cases &  21 & 5 & 18 & 145\\
			\hline
			Issue area &  Judicial Power
			& Federalism & Federal taxation & Total \\ 
			Number of cases & 133 & 57 & 20 & 893 \\
			\hline
	\end{tabular}}	
\end{table}

The Supreme Court Database has classified each case into one of 14 issue areas, such as criminal procedure and civil rights. Table~\ref{tab:case} presents the number of cases associated with each issue area; cases were only heard in 11 out of 14 areas during the Second Rehnquist Court. The database has also classified the decisions for each case as either liberal or conservative.  For each case, each justice has an outcome, $Y$, which denotes whether the justice's personal opinion was liberal ($Y=$+1) or conservative ($Y=-1$); the ruling in the case is liberal if at least 5 of the justices form liberal opinions and conservative otherwise. 

Using data on the case issue area and binary outcomes of each justice for $n=893$ cases, we examined the effect of issue area on whether the individual justices reached conservative or liberal opinions, whether the court reached a conservative versus a liberal ruling, and whether the decision was unanimous or divided. (Only group-level exposure variables are available in the database; in the next section we simulate individual-level treatments to illustrate the use of chain graph models in more general settings.) There is strong evidence (including self-report by the justices) that the Court works hard to come to unanimous decisions, but 5-to-4 decisions are frequent~\citep{sunstein2014unanimity, riggs1992every}, and we found that there is an effect of issue area on this outcome.  There is also considerable academic interest in each justice's personal orientation~\citep{songer1996not, tate1981personal}. During the Rehnquist court, 56$\%$ of the decisions were conservative. Clarence Thomas was the most conservative justice, signing the conservative opinion in 72$\%$ of cases, while Ruth Bader Ginsburg was the most liberal, signing the liberal opinion in 60$\%$ of cases.  We found that issue area had a strong effect on both individual outcomes and on overall court decisions, which is consistent with literature on the effect of issue areas on the ideology of each justice and on the final decision of Supreme Court~\citep{tate1981personal, lu2011understanding}.  

We considered the effects of (i) criminal procedure, (ii) civil rights, (iii) economic activity, and (iv) judicial power on conservative vs liberal opinions. In order to illustrate our method for binary treatment variables, and because considering a higher dimensional treatment variable would require a larger sample size, we fit separate chain graph models for each issue area, coded $a=1$ for the issue area of interest and $a=0$ for all other issue areas. Although issue area is not manipulable in practice, these causal effects are of explanatory interest: the effect of a hypothetical intervention that holds all relevant features of a case constant while changing the issue area can elucidate the role that issue area plays in the collective decision-making of the justices, even if it cannot inform policy.   

\begin{figure}[ht]
	\begin{centering}
		\begin{tikzpicture}[>=stealth, node distance=3.0cm]
		
		\tikzstyle{format} = [very thick, rectangle, minimum size=5.0mm,
		inner sep= 2pt]
		\tikzstyle{square} = [very thick, rectangle, draw]
		
		\draw[thick, rounded corners] (-3,3.5) rectangle (7,-7);

		\begin{scope}[xshift=0.0cm]
		\path[very thick, ->]                
		node[format, xshift=2cm, col10, fill = col4] (1) {$Y_{\mbox{Rehnquist}}$}
		node[format, right of=1, col10, fill = col7] (9) {$Y_{\mbox{Breyer}}$}
		node[format, above of=9, col10, fill = col7] (2) {$Y_{\mbox{Stevens}}$}
		node[format, left of = 1, col10, fill = col4] (3) {$Y_{\mbox{O'Connor}}$}
		
		node[format, xshift = -3cm, yshift = -2cm] (l) {}
		node[left of=l] (a) {$A_{\mbox{case : Judicial Power}}$}
		node[ below of=a] (x) {}	
		node[ xshift=0.4cm, left of=3] (m) {}
		
		node[format, below of=3, col10, fill = col4] (5) {$Y_{\mbox{Kennedy}}$}
		node[format, below of=5, col10, fill = col4] (4) {$Y_{\mbox{Scalia}}$}
		
		node[format, below of=9, col10, fill = col7] (6) {$Y_{\mbox{Souter}}$}
		node[format, above of=3, col10, fill = col4] (7) {$Y_{\mbox{Thomas}}$}
		node[format, below of=6, col10, fill= col7] (8) {$Y_{\mbox{Ginsburg}}$}

		(a) edge[->, black] (l)
		
		(1) edge[-] (3) 
		(1) edge[-] (4) 
		(1) edge[-] (5) 
		(1) edge[-] (7) 

		(2) edge[-, bend left] (6) 
		(2) edge[-, bend left] (8) 
		(2) edge[-] (9) 
		
		(3) edge[-] (5) 
		(3) edge[-] (6) 
		(3) edge[-, bend left] (9) 
		
		(4) edge[-] (5) 
		(4) edge[-, bend left] (7) 
		
		(5) edge[-] (6) 
		(5) edge[-, bend left] (7) 
		(5) edge[-] (8) 
		
		(6) edge[-] (8) 
		(6) edge[-] (9) 

		(8) edge[-, bend left] (9) 
		;
		\end{scope}
		
		\end{tikzpicture} 
		\par\end{centering}
	\caption{\label{fig:supreme_network_judicial_unweighted} The estimated underlying network among the nine justices for cases involving judicial power. The color of each node indicates whether the justice is commonly considered to be liberal (blue) or conservative (red). 
	}
\end{figure}
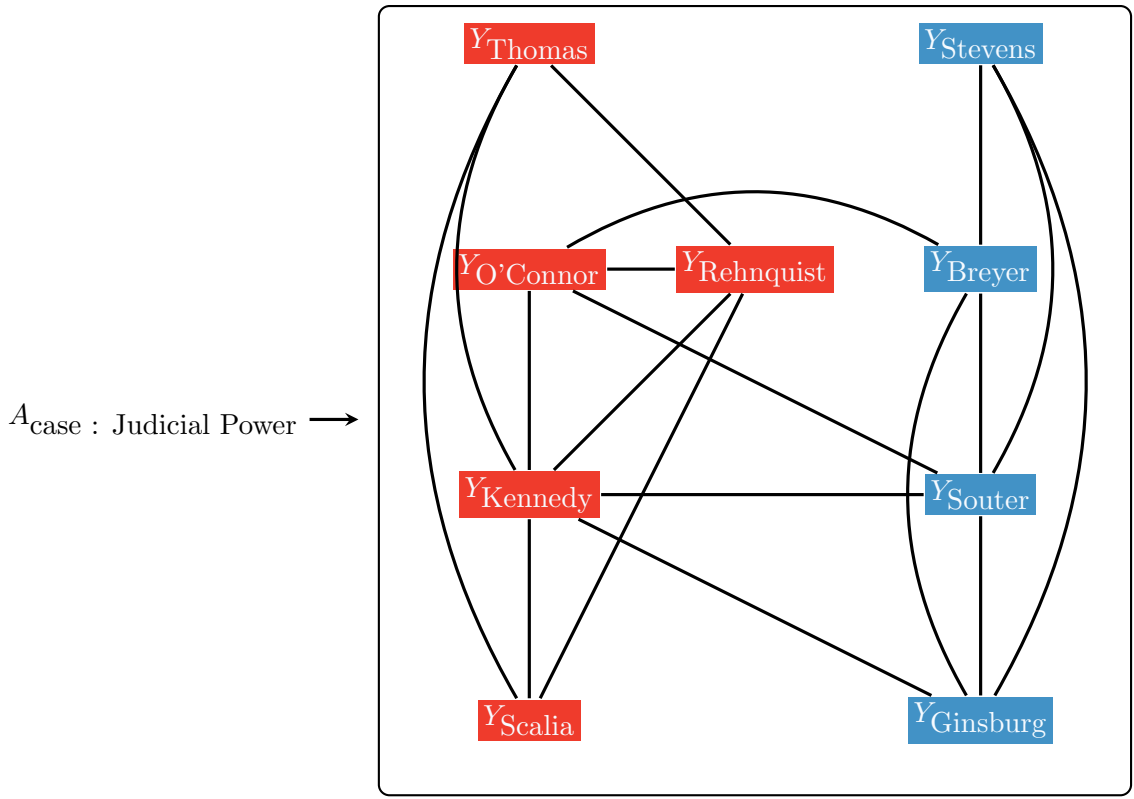

So far, we have assumed that the underlying network according to which individuals interact is known. In this case, however, we have only  anecdotal evidence about the relationships among the justices.  Therefore, as a first step, we estimated the undirected component of the chain graph model using data on the justices' opinions for a particular issue area.  Estimation was implemented with the \texttt{XMRF} \texttt{R} package \citep{wan2016xmrf}, which used gradient descent to maximize penalized node-conditional likelihoods and fit an exponential family Markov random field model to the data.  Incorporating uncertainty about the underlying network into causal inference is beyond the scope of this paper, but is a problem on which we are actively working.

To complete the chain graph of interest, we added a single treatment variable, i.e. issue area, that jointly affects each justice's outcome. Our causal chain graph models and resulting analyses assume  that there are no confounders of the relationship between issue area and justices' decisions.  We think this is a plausible assumption, but it is possible that there could be (unmeasured) confounding, for example due to the court's selection of cases to hear, or due to time trends in both issue areas and justices' behavior. We demonstrate how to use chain graph models with individual-level treatments and confounders in Section \ref{ssec:supreme_sim}.

The resulting chain graph for the judicial power issue area is displayed in Figure~\ref{fig:supreme_network_judicial_unweighted}. 
We found that justices interact with one another based on their shared liberal or conservative leanings, as would be expected, but also across that divide, in some cases through relationships supported by anecdotal evidence. For example, 
the tie between Breyer and O'Connor could be explained by their social connections\footnote{\url{http://blogs.findlaw.com/supreme_court/2017/03/supreme-court-shutters-justice-oconnors-workout-class.html}} or their shared views on judicial independence\footnote{\url{http://www.pbs.org/newshour/bb/law-july-dec06-independence_09-26/}}. 

Separately for each of the four issue areas, we estimated the parameters of the following chain graph model:
\begin{equation}
\label{eq:supreme}
p\big( \mathbf{Y}(a) = (y_{1}, y_{2}, \ldots, y_{9})  \big) = \frac{1}{Z(a)} \exp \left\{\sum\limits_{i=1}^{9} h_{i}y_{i} + \sum\limits_{i,j=1, e_{ij} = 1 }^{9} k_{ij} y_{i} y_{j}  +  \sum\limits_{i=1}^{9} \gamma_{i} a y_{i}  \right\},
\end{equation}
where $\mathbf{Y}(a)$ is the vector of counterfactual outcomes under $A=a$, $e_{ij} = 1$ implies justice $i$ and $j$ share an undirected edge in the chain graph, and  $Z(a)$ is a normalizing constant. 
The parameters $h$, $k$, and $\gamma$ all relate to conditional log odds (in this particular model they must first be multiplied by $2$, since $Y$ is coded $1$ or $-1$ instead of $1$ or $0$).  The parameter $h_{i}$ represents the conservative or liberal leaning of Justice $i$, with a positive $h_{i}$ indicating bias towards liberal opinions.  Specifically, $2h_{i}$ is the conditional log odds of $y_{i} = 1$ given that all the parameters associated with relational dependence of $y_{i}$ ($\{ k_{ij}: e_{ij} = 1 \}$) and with the treatment ($\gamma_{i}$) are zero.  The interaction parameter $k_{ij}$ captures the tendency of Justices $i$ and $j$ to agree beyond what can be explained by their individual leanings, with a positive $k_{ij}$ indicating tendency to agree while a negative $k_{ij}$ indicates tendency to disagree ($2 k_{ij}$ is the conditional log odds ratio for the association between $Y_{i}$ and $Y_{j}$).  The parameter $\gamma_i$ gives the causal effect of issue area $A$ on Justice $i$'s opinions, with positive $\gamma_{i}$ indicating tendency toward liberal opinions above and beyond what can be explained by the Justice's independent leaning or by the interactions with the other justices ($2\gamma_{i}$ is the conditional log odds ratio for $Y_{i} = 1$ when $A = 1$ compared to $A=0$).  In principle, higher order interactions could be added to the model to capture tendencies of larger groups justices to agree or disagree beyond what the pairwise interactions explain, but there was no empirical evidence to support the inclusion of these interactions (see Supporting Material). We bootstrapped the standard errors in order to calculate 95$\%$ confidence intervals for the parameters of interest, with $nb = 500$ bootstrap samples for each model.

\begin{figure}[ht]
	\begin{centering}
		\begin{tikzpicture}[>=stealth, node distance=3.0cm]
		
		\tikzstyle{format} = [very thick, rectangle, minimum size=5.0mm,
		inner sep= 2pt]
		\tikzstyle{square} = [very thick, rectangle, draw]
		
		\draw[thick, rounded corners] (-3,3.5) rectangle (7,-7);
		
		\begin{scope}[xshift=0.0cm]
		\path[very thick, ->]                
		node[format, xshift=2cm, col10, fill = col2] (1) {$Y_{\mbox{Rehnquist}}$}
		node[format, right of=1, col10, fill = col6] (9) {$Y_{\mbox{Breyer}}$}
		node[format, above of= 9, col10, fill = col9] (2) {$Y_{\mbox{Stevens}}$}
		node[format, left of = 1, col10, fill = col2] (3) {$Y_{\mbox{O'Connor}}$}
		
		node[format, xshift = -3cm, yshift = -2cm] (l) {}
		node[left of=l] (a) {$A_{\mbox{case : Judicial Power}}$}
		node[ below of=a] (x) {}	
		node[ xshift=0.4cm, left of=3] (m) {}
		
		node[format, below of=3, col10, fill = col5] (5) {$Y_{\mbox{Kennedy}}$}
		node[format, below of=5, col10, fill = col4] (4) {$Y_{\mbox{Scalia}}$}
		
		node[format, below of=9, col10, fill = col8] (6) {$Y_{\mbox{Souter}}$}
		node[format, above of=3, col10, fill = col1] (7) {$Y_{\mbox{Thomas}}$}
		node[format, below of=6, col10, fill= col7] (8) {$Y_{\mbox{Ginsburg}}$}

		
		(1) edge[-, line width = 1.2308982] (3) 
		(1) edge[-, line width = 0.6905832] (4) 
		(1) edge[-, line width = 1.0626196] (5) 
		(1) edge[-, line width = 0.7584730 ] (7) 

		(2) edge[-, bend left, line width = 0.6888480 ] (6) 
		(2) edge[-, bend left, line width = 0.7816874 ] (8) 
		(2) edge[-, line width = 0.7099157]  (9) 
		
		(3) edge[-, line width = 0.6298417] (5) 
		(3) edge[-, line width = 0.6776570] (6) 
		(3) edge[-, bend left, line width = 1.0187295] (9) 
		
		(4) edge[-, line width = 0.5025447] (5) 
		(4) edge[-, bend left, line width = 2.0160801 ] (7) 
		
		(5) edge[-, line width = 0.5637078 ] (6) 
		(5) edge[-, bend left, line width = 0.5575171 ] (7) 
		(5) edge[-, line width = 0.5045031] (8) 
		
		(6) edge[-, line width = 1.2431446 ] (8) 
		(6) edge[-, line width = 0.6058508] (9) 

		(8) edge[-, bend left, line width = 1.0313114 ] (9) 
		
		(a) edge[->, blue, line width = 0.32402724] (1)
		(a) edge[->, bend left, red, line width = 2.22082220] (2)
		(a) edge[->, blue, line width = 0.84974616] (3)
		(a) edge[->, red, line width = 1.22317942] (4)
		(a) edge[->, blue, line width = 0.62502044] (5)
		(a) edge[->, red, line width = 0.26354494] (6)
		(a) edge[->, blue, line width = 1.48718375] (7)
		(a) edge[->, red, line width = 1.14375875] (8)
		(a) edge[->, blue, line width = 0.05223488] (9)
		;
		\end{scope}
		
		\end{tikzpicture} 
		\par\end{centering}
	\caption{\label{fig:supreme_network_judicial_weighted} The color of each node represents the justice's tendency towards liberal or conservative decisions, estimated by  $\hat{h}_{i}$, with darker red representing more conservative and darker blue more liberal decisions. The width of the undirected edges represents the strength of pairwise interactions between justices, estimated by $\hat{k}_{ij}$.  The color and width of the directed edges represent the direction and magnitude of the individual-level causal effects of an intervention on whether or not a case concerns judicial power, estimated by $\hat{\gamma}_i$, with red for negative and blue for positive effects.}
\end{figure}
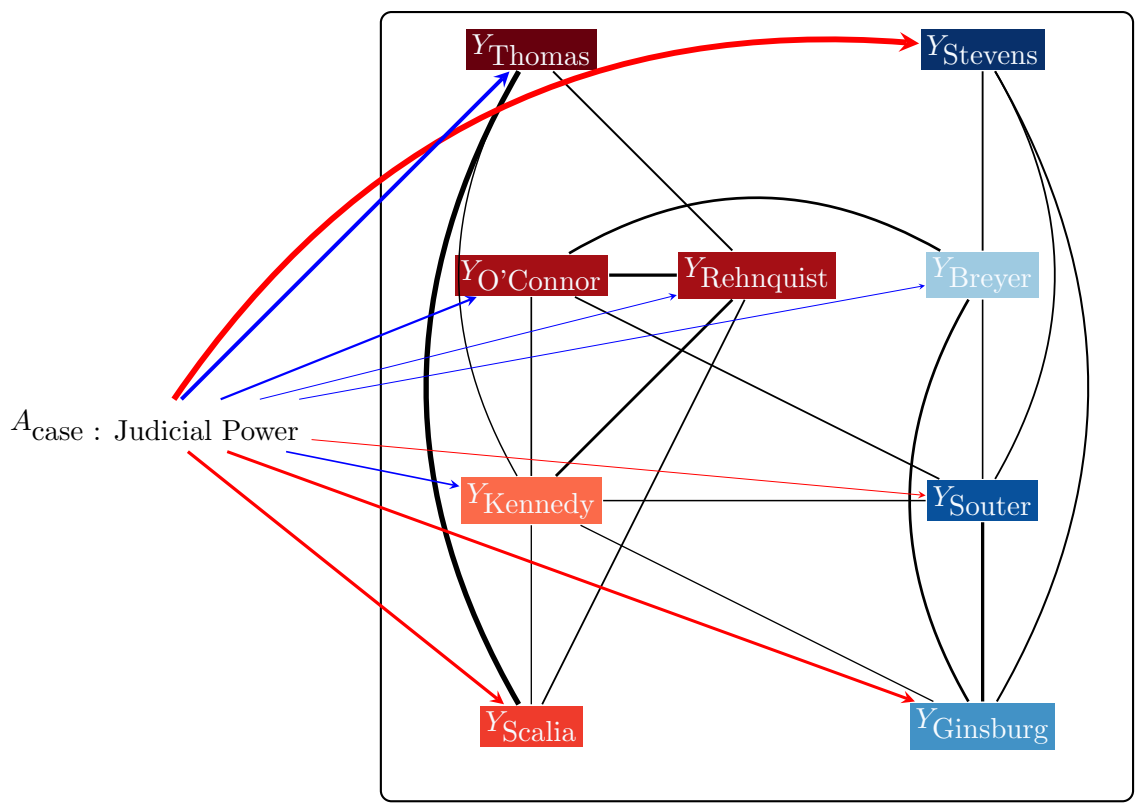

Figure~\ref{fig:supreme_network_judicial_weighted} illustrates the estimated parameter values for model \ref{eq:supreme}, when $A$ is an indicator of judicial power versus other issue areas.  The shade of the node reflects the estimated main effect ($\hat{h}_{i}$), the width of the undirected edges reflects the magnitude and sign of the estimated interaction between justices $i$ and $j$ ($\hat{k}_{ij}$), and the width and color of a directed edges reflects the estimated effect of $A$ on Justice $i$'s decisions ($\hat{\gamma}_{i}$), when $A$ represents judicial power versus other issue areas. 

Using the model given in Equation \ref{eq:supreme}, we estimated the causal effects of each issue area on the majority-based decisions of the nine justices. As an example, the causal effect (on the risk difference scale) of the judicial power issue area on the probability of a unanimous liberal decision (that is, $\mathbf{Y}=\mathbf{1}$) is derived below: 
\begin{eqnarray*}
	&& p(\mathbf{Y}(1) = \mathbf{1} ) - p(\mathbf{Y} (0)= \mathbf{1} ) \\ & = & \frac{1}{Z(1)} \exp \left\{ \sum\limits_{i=1}^{9} h_{i} + \sum\limits_{i,j=1, e_{ij} = 1}^{9} k_{ij} + \sum\limits_{i=1}^{9} \gamma_{i} \right\} - \frac{1}{Z(0)} \exp \left\{ \sum\limits_{i=1}^{9} h_{i} + \sum\limits_{i,j=1, e_{ij} = 1}^{9} k_{ij}  \right\}	\\
	& = & \exp \left\{ \sum\limits_{i=1}^{9} h_{i} + \sum\limits_{i,j = 1, e_{ij} = 1}^{9} k_{ij} \right\} \left\{ \exp( \sum\limits_{i=1}^{9} \gamma_{i} ) / Z(1) - 1 / Z(0) \right\} 
\end{eqnarray*}	
When $\sum\limits_{i=1}^{9} \gamma_{i}$ = 0, i.e. when the sum of the log of odds ratios is equal to zero, the causal effect is also equal to zero.  This causal effect, which aggregates across all of the justices, depends not only on the $\gamma_{i}$ parameters but also on the main effects ($h_{i}$) and interactions ($k_{ij}$). We can similarly derive the effect of issue area on the probability of a unanimous conservative decision, that is $\mathbf{Y} = \mathbf{0}$, or on split 5-4 decisions, where $\sum{Y_i}=5$ represents a 5-to-4 liberal decision and $\sum{Y_i}=4$ represents a 5-to-4 conservative decision.  We could also derive causal effect estimates on the risk ratio or odds ratio scale.
We found that judicial power had the strongest causal effect on the probability of unanimous conservative decisions (Table~\ref{tab:judicial}). Economic activity caused an increase in the probability of liberal unanimous decisions (Supporting Material Table 4). Criminal procedures and civil rights both caused an increase in the probability of 5-to-4 conservative decisions (Supporting Material Tables 2 and 3).

\begin{table}
	\caption{\label{tab:judicial} The estimated potential outcomes and causal effects (standard errors in the parentheses) comparing judicial power ($a=1$) to other issue areas ($a=0$), where the outcomes of interest are the probabilities of unanimous decisions.}	
	\centering
	\fbox{%
		\begin{tabular}{*{3}{l}}
			\hline
			 & $ p(\mathbf{Y}(a) = \mathbf{0} ) $, probability of a & $ p(\mathbf{Y}(a) = \mathbf{1} ) $, probability of a  \\
			  & unanimous conservative decision &  unanimous liberal decision\\ 
			\hline
			Judicial power ($a=1$) & 0.33  ( 0.03  )  &   0.16  ( 0.02  )  \\ 
			Other issues ($a=0$ )& 0.20  ( 0.01  )  &    0.17  ( 0.01  )  \\ 
			\hline
			Causal Effect & 0.13  ( 0.03  )  &   -0.01  ( 0.03  )  \\ 
			\hline		
	\end{tabular}}

\end{table}

\section{Simulations with individual-level treatments and covariates}
\label{ssec:supreme_sim}

In the application above, we considered a single treatment variable that jointly affects the justices' opinions, and we assumed no confounding. 
To illustrate how chain graphs can be used to estimate causal effects with individual-level treatments and with confounders, we simulated data from the undirected component of the graph in Figure \ref{fig:supreme_network_judicial_unweighted} with the addition of individual-level treatments $A$, and individual level binary confounders $C$ that are dependent across justices and have direct casual effects on $A$ and $Y$. Treatment $A_i \in \{ 0, 1\}$ nudges Justice $i$ towards a liberal ($A=1$) or conservative ($A=0$) decision.  For an initial simulation, we set the main effect and pairwise interaction parameters to their estimates from a log-linear model fit to the actual Supreme Court data.  We then varied the magnitude of the main effects and two-way interaction terms by multiplying those parameters by $\alpha, \beta \in \{ 0.5 , 1 \}$, respectively.  
Specifically, we generated the outcomes $\mathbf{Y}$ from the chain graph model below, where $h_i$ and $k_{ij}$ were estimated from the real data, and we varied $\alpha$ and $\beta$ across simulation settings: 
\begin{equation}
\begin{split}
	\label{eq:simmodel}
	 p\big( \mathbf{Y} & = (y_{1}, y_{2}, \ldots, y_{9}) | \mathbf{A} = \mathbf{a}, \mathbf{C} = \mathbf{c} \big)\\ & = \frac{1}{Z(\mathbf{a}, \mathbf{c}) }    \exp \left\{ \alpha\sum\limits_{i=1}^{9} h_{i} y_{i} + \beta  \sum\limits_{i,j=1, e_{ij} = 1}^{9}  k_{ij} y_{i} y_{j}  +  \gamma \sum\limits_{i=1}^{9}  a_{i} y_{i} + \kappa \sum\limits_{i=1}^{9} c_{i}y_{i}  \right\}.
\end{split}
\end{equation}
For each combination of $\alpha$ and $\beta$, we generated $500$ simulated data sets from the chain graph model, each of which used Gibbs sampling to produce $2000$ observations of $(\mathbf{Y}, \mathbf{A}, \mathbf{C})$. Additional details of the simulation are in the Supporting Material. 

Causal effects comparing two different treatment vectors,  $\mathbf{a}_{1} = (a_{1,1}, a_{2,1}, \ldots, a_{1,9})$ and $\mathbf{a}_{0} = (a_{0, 1}, a_{0, 2}, \ldots, a_{0, 9})$, are functions of the parameters of the chain graph model in Equation \ref{eq:simmodel}.  For example, causal effect on the probability of unanimous liberal decisions (on the risk difference scale) is given by
\begin{eqnarray*}
 &&p(\mathbf{Y}( \mathbf{a}_{1}) = \mathbf{1}) - p(\mathbf{Y}(\mathbf{a}_{0}) = \mathbf{1}) \\
 & = &  \sum\limits_{ \mathbf{C}} \left\{p(\mathbf{Y}  =  \mathbf{1} | \mathbf{A} = \mathbf{a}_{1}, \mathbf{C} = \mathbf{c} ) - p(\mathbf{Y} = \mathbf{1} | \mathbf{A} = \mathbf{a}_{0}, \mathbf{C} = \mathbf{c} )  \right\} p(\mathbf{C} = \mathbf{c}) \\ 
 & = &  \sum\limits_{ \mathbf{C} } \exp \left\{  \alpha \sum\limits_{i=1}^{9} \hat{h}_{i} + \beta \sum\limits_{i,j=1, e_{ij} = 1}^{9} \hat{k}_{ij} + \kappa \sum\limits_{i=1}^{9} c_{i}  \right\}  \left\{ \frac{\gamma \sum\limits_{i=1}^{9}a_{1,i} }{Z(\mathbf{a}_{1}, \mathbf{c})} - \frac{\gamma \sum\limits_{i=1}^{9} a_{0,i}}{Z(\mathbf{a}_{0}, \mathbf{c})} \right\} p(\mathbf{c}).
\end{eqnarray*}

\begin{table}
	\caption{\label{tab:gibbs_true} Probability of having unanimous liberal decision ($\sum \mathbf{y} = 9$), unanimous conservative decision ($\sum \mathbf{y} = 0 $), five-liberal votes ($\sum \mathbf{y} = 5$), and five-conservative votes ($\sum \mathbf{y} = 4$) under six different treatment assignments. 
	}
	\centering
	\resizebox{0.8\textwidth}{!}{\begin{tabular}{l||l|l}
			\hline
			Intervened justice ($\mathbf{a}$) & $P( \mathbf{Y} (\mathbf{a}) = \mathbf{y};  \sum \mathbf{y} = 9  )$ &$P( \mathbf{Y} (\mathbf{a}) = \mathbf{y};  \sum \mathbf{y} = 0  )$   \\ 		
			\hline
			O'Connor, Scalia, Kennedy, Thomas & 0.37 & 0.11 \\ 
			Stevens, Souter, Ginsburg, Breyer & 0.21 & 0.06  \\ 
			Rehnquist& 0.16 & 0.23  \\ 
			Thomas  & 0.17 & 0.24 \\ 
			Stevens& 0.13 & 0.19 \\ 
			Scalia  & 0.16 & 0.23  \\ 
			\hline	
			Intervened justice ($\mathbf{a}$) &  $P( \mathbf{Y} (\mathbf{a}) = \mathbf{y};  \sum \mathbf{y} = 5  )$ & $P( \mathbf{Y} (\mathbf{a}) = \mathbf{y};  \sum \mathbf{y} = 4  )$  \\ 		
			\hline
			O'Connor, Scalia, Kennedy, Thomas & 0.06 & 0.06 \\ 
			Stevens, Souter, Ginsburg, Breyer  & 0.13 & 0.19 \\ 
			Rehnquist & 0.07 & 0.10  \\ 
			Thomas& 0.07 & 0.10 \\ 
			Stevens & 0.09 & 0.13 \\ 
			Scalia  & 0.07 & 0.10 \\ 
			\hline						
	\end{tabular}}
\end{table}

Table~\ref{tab:gibbs_true} presents the true probability of four different counterfactual outcomes  when $\alpha = \beta = 1$, under six different treatment assignments (treating four conservative justices; treating four liberal justices; treating chief justice Rehnquist; treating Justice Thomas; treating Justice Stevens; treating Justice Scalia). The table shows that in our simulated data, treating the four conservative justices results in a higher probability of unanimous liberal decisions (0.37) than treating the four liberal justices (0.21). Similarly, treating the most liberal justice (Stevens) has the smallest effect on the probability of unanimous liberal decisions compared to treating Justices Rehnquist, Thomas, or Scalia. Treating Justice Stevens has the greatest impact on the probabilities of 5-to-4 or 4-to-5 decisions.Treating the four liberal justices together results in relatively high probability of 4(liberal)-to-5(conservative) decisions (0.19).  When we estimated each of these probabilities in 500 simulations, the average absolute bias ranged from $0.002$ to $0.055$ and the average standard error ranged from $0.003$ to $0.022$; full results are in the Appendix.

As an example, Figure~\ref{fig:supremesim} shows the true and estimated potential outcomes--in this case probability of unanimous decisions--under two different treatment vectors.  Treatment  $\mathbf{a}_{1}$ treats (nudges towards a liberal decision) the four most conservative justices (Justices O'Connor, Scalia, Kennedy, and Thomas), and treatment $\mathbf{a}_{0}$ treats the four most liberal justices (Justices Stevens, Souter, Ginsberg, and Breyer).  We found that treating the conservative justices had a significant positive effect on the probability of a unanimous liberal decision compared to treating the liberal justices.   On the other hand, there was no significant causal effect on the probability of unanimous conservative decisions. 
In the Supporting Materials, we also present similar results under different simulated magnitudes of the main effect parameters for each justice (controlled by $\alpha$) and pairwise interactions for connected justices (controlled by $\beta$), and for different treatment vectors.
All the code and accompanying data can be found in \url{https://github.com/youjin1207/Chain}.

\begin{figure}
	\centering
	\makebox{
	\includegraphics[width=0.7\textwidth]{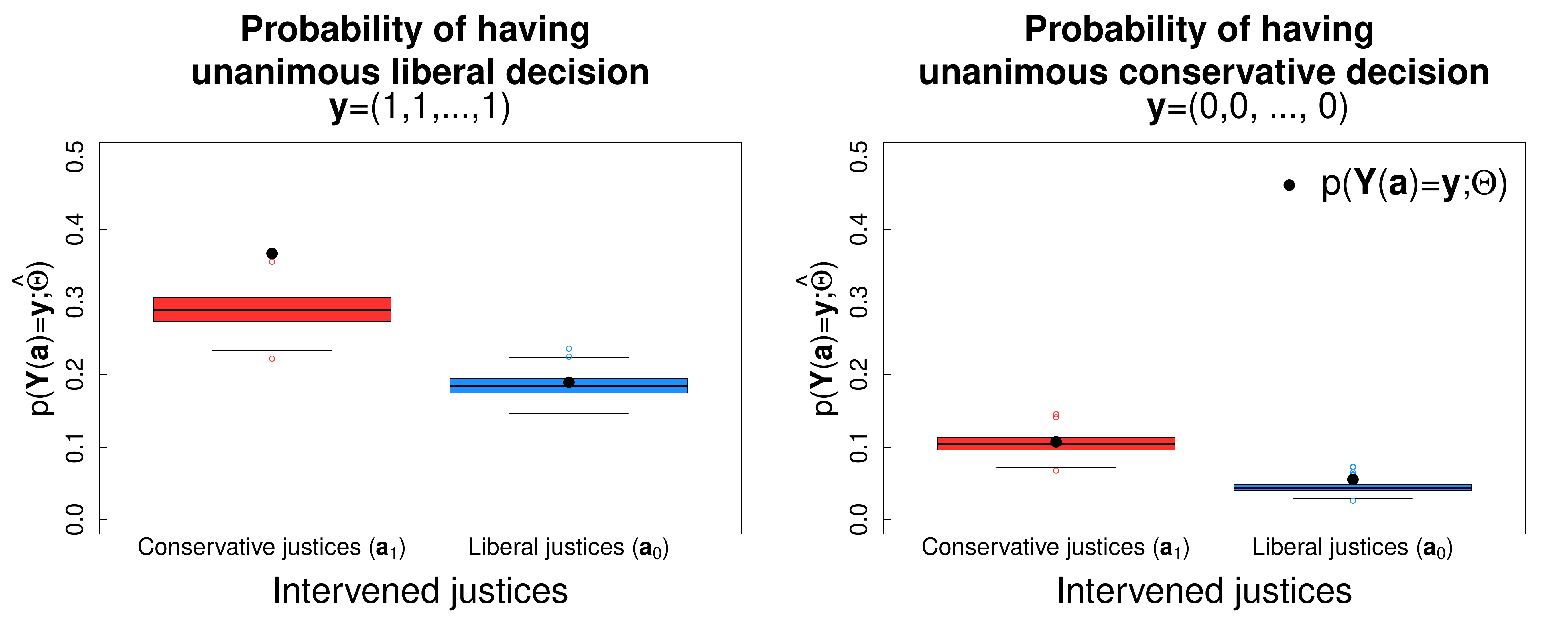}
	}
	\caption{\label{fig:supremesim} The effects of intervening on the 4 liberal ($\mathbf{A} = \mathbf{a}_{1}$) or the 4 conservative justices ($\mathbf{A} = \mathbf{a}_{0}$) on the probabilities of unanimous liberal ($\mathbf{y} = (y_{1}, y_{2}, \ldots, y_{9}) = \mathbf{1}_{9}$) and unanimous conservative decisions ($\mathbf{y} = (y_{1}, y_{2}, \ldots, y_{9}) = \mathbf{0}_{9}$). }
\end{figure}
\

	

\section{Conclusion and next steps} \label{sec:conclusion}

We have described a chain graph model for outcomes associated with nodes in a social network, with dependence along network ties induced by social interactions, contagion, or interference.  Although our Supreme Court example afforded multiple, ostensibly i.i.d. observations from the same chain graph, statistical inference for a single realization of a large chain graph was developed in \cite{tchetgen2017auto}. The chain graph model can only represent the true data-generating distribution if the outcomes are in very specific kinds of equilibria, which may be plausible if the outcomes represent collective beliefs or decisions (as in the Supreme Court example) but are often implausible.  However, we showed that data generated from a causal DAG may be well-approximated by a chain graph model, clarifying the conditions under which it may be reasonable to use of these models in the literature on Ising models for collective behavior, undirected models for social networks, and chain graph models for causal inference in social networks. This approximation has two major limitations: first, it requires that the outcome evolve slowly over time, and second, it requires that the only source of dependence across nodes in the network be due to the causal effects from one node's outcome to another.  This rules out \emph{latent variable dependence}, where outcomes from nodes that are connected or close in the social network may be dependent due to shared environment, genetics, or other characteristics \cite{ogburn2017causal, ogburn2018}.  Future work is needed to develop tractable statistical models that can handle these more general kinds of dependence, and to test these models against real-world data.

\section*{Acknowledgements}
 ELO and YL were supported by ONR grant N000141512343. ELO, YL, and IS were supported by ONR grant N000141812760.

\bibliographystyle{chicago}
\bibliography{references}

\end{document}